\documentclass[12pt]{article} \usepackage{dina4p} \usepackage{my}
\usepackage{epsfig,amsmath} \usepackage{graphics}
 
\def\lsim{\mathrel{\rlap{\lower4pt\hbox{\hskip1pt$\sim$}}
    \raise1pt\hbox{$<$}}}                
\def\gsim{\mathrel{\rlap{\lower4pt\hbox{\hskip1pt$\sim$}}
    \raise1pt\hbox{$>$}}}                

\newcommand{\as}{\alpha_\mathrm{s}}

\newcommand{\Pmax}{\bar{q}}
\newcommand{\kt}{k_{t}}
\newcommand{\ktp}{k_{t}^{\prime}}

\newcommand{\CCFM}{CCFMa,CCFMb,CCFMc,CCFMd}
\newcommand{\BFKL}{BFKLa,BFKLb,BFKLc}

\newcommand{\PYTHIAMC}{Jetsetc}

\newcommand{\DGLAP}{DGLAPa,DGLAPb,DGLAPc,DGLAPd}
\def\CASCADE{{\sc Cascade}}

\def\PYTHIA{{\sc Pythia}}

\begin{document}
\begin{flushright}
 DESY 01-136\\
 LUNFD6/(NFFL--7203) 2001 \\
 hep-ph/0110034\\
\end{flushright}

\vspace*{10mm}
\begin{center}  \begin{Large} \begin{bf}
{\renewcommand{\thefootnote}{\fnsymbol{footnote}}
Heavy Quark production at the TEVATRON and HERA 
using {\boldmath $k_t$} - factorization with CCFM evolution
 }\\
  \end{bf}  \end{Large}
  \vspace*{5mm}
  \begin{large}
 H. Jung \vspace{0.3cm}\\
  \end{large}
{\small
Physics Department, Lund University, Box 118, S-221~00 Lund, Sweden }
\end{center}
\setcounter{footnote}{0}
\begin{quotation}
\noindent
{\bf Abstract:} The application of $k_t$-factorization supplemented with the
CCFM small-$x$ evolution equation to heavy quark production  
 at the TEVATRON and at HERA is discussed.
The $b\bar{b}$
production cross sections at the TEVATRON
can be consistently described using the $k_t$-factorization 
formalism together with the unintegrated gluon density obtained
within the CCFM evolution approach from a fit to HERA $F_2$ data. 
Special attention is drawn to the comparison with measured 
visible cross sections, which are compared to the hadron level
Monte Carlo generator \CASCADE .
\end{quotation}
\section{Introduction}
The calculation of inclusive quantities, like the structure function
$F_2(x,Q^2)$ at HERA, performed in NLO QCD is in perfect agreement with the
measurements.  
However, Catani argues, that the NLO approach,
although phenomenologically successful for $F_2(x,Q^2)$, is
not fully satisfactory from a theoretical viewpoint, because 
{\it ``the truncation of the
splitting functions at a fixed perturbative order is equivalent to assuming that
the dominant dynamical mechanism leading to scaling violations is the evolution
of parton cascades with strongly ordered transverse 
momenta"}~\cite{catani-feb2000}.
As soon as exclusive quantities like jet or heavy quark
production  are investigated, the agreement between NLO coefficient functions
convoluted with NLO DGLAP~\cite{\DGLAP}
 parton densities and the data is not at all
satisfactory: large so-called $K$-factors (normalization factors)
\cite{CDF_bbar,D0_bbar,H1_bbar,ZEUS_bbar}
 are needed
to bring the NLO calculations close to the data ($K \sim 2-4$ for bottom
production at the TEVATRON),
indicating that in the calculations a significant
part of the cross section is still missing.
\par 
At small $x$ the structure function $F_2(x,Q^2)$ is proportional
to the sea quark density, and the sea-quarks are driven via the DGLAP evolution
equations by the gluon density. Standard QCD fits 
determine the parameters of
the initial parton distributions at a starting scale $Q_0$. With help of the
DGLAP evolution equations these parton distributions are then evolved to any
other scale $Q^2$, with the splitting functions still truncated at fixed
$O(\alpha_s)$ (LO) or $O(\alpha_s ^2)$ (NLO).
Any physics process in the fixed order scheme is then calculated via 
collinear factorization into the coefficient functions
$C^a(\frac{x}{z})$ and collinear (independent of $k_t$) 
parton density functions:
$f_a(z,Q^2)$:
\begin{equation}
\sigma = \sigma_0 \int \frac{dz}{z} C^a(\frac{x}{z}) f_a(z,Q^2)
\label{collinear-factorisation}
\end{equation}
At large energies (small $x$) the evolution of parton densities proceeds over a large
region in rapidity $\Delta y \sim \log(1/x)$ and effects of finite transverse
momenta of the partons may become increasingly important.
Cross sections can then be $k_t$ - factorized~\cite{CCH}
into an off-shell ($k_t$ dependent) partonic cross section
$\hat{\sigma}(\frac{x}{z},k_t) $
and a $k_t$ - unintegrated parton density function 
${\cal F}(z,k_t)$:
\begin{equation}
 \sigma  = \int 
\frac{dz}{z} d^2k_t \hat{\sigma}(\frac{x}{z},k_t) {\cal F}(z,k_t)
\label{kt-factorisation}
\end{equation}
The unintegrated gluon density ${\cal F}(z,k_t)$ is 
described by the BFKL~\cite{\BFKL}
 evolution equation in the region of asymptotically large energies (small $x$). 
 An appropriate description valid for
both small and large $x$ is given by the CCFM evolution
equation~\cite{\CCFM}, resulting in an unintegrated gluon density 
${\cal A} (x,\kt,\Pmax ) $, which is a function also of the 
additional evolution scale $\Pmax $ described below.
\par
In \cite{catani-dis96,catani-feb2000} Catani argues that
by explicitly carrying out the $k_t$ integration in eq.(\ref{kt-factorisation})
one can obtain a form fully consistent with collinear factorization: the
coefficient functions and also the DGLAP splitting functions leading to 
$f_a(z,Q^2)$ are no longer evaluated in fixed order perturbation theory but
supplemented with the all-order resummation of the $\as \log 1/x$ contribution  
at small $x$. 
\par
The application of  $k_t$ - factorization to heavy quark hadroproduction is
discussed in detail in~\cite{LRSS,RSS2001,RSS2000}.
In the present  paper  bottom production at the TEVATRON and at HERA 
is  investigated using the $k_t$ - factorization approach.
The unintegrated gluon density has been obtained 
previously in \cite{jung_salam_2000} from a
CCFM fit to the HERA structure function $F_2(x,Q^2)$. All free
parameters are thus fixed and absolute predictions for bottom production can be
made. The universality of the unintegrated CCFM gluon distribution 
will be tested by the application to TEVATRON and HERA results.
\par
The basic features of the CCFM evolution
equation are recalled and 
the unintegrated gluon density is investigated. Then 
the calculations for $b \bar{b}$ production at the TEVATRON is presented as well
as calculations of the visible cross section for $b\bar{b}$
production at HERA.
\section{The CCFM evolution equation}
\label{sec:CCFMEquation}

\begin{figure}
\begin{center} 
\epsfig{figure=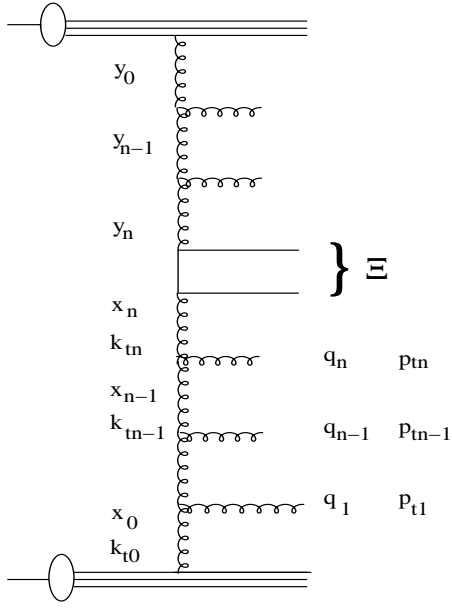,width=6cm,height=8cm}
\end{center} 
\caption{{\it Kinematic variables for multi-gluon emission. The $t$-channel gluon
four-vectors are given by $k_i$ and the gluons emitted in the initial state
cascade have four-vectors $p_i$. The minimum (maximum)
 angle for any emission is obtained
from the quark box, as indicated with $\Xi$.  
\label{CCFM_variables} }} 
\end{figure}
A solution of the CCFM evolution equation, which properly describes the
inclusive structure function $F_2(x,Q^2)$ and also typical small $x$ final
state processes at HERA has been presented in detail in ~\cite{jung_salam_2000}. 
Figure~\ref{CCFM_variables} illustrates the pattern of QCD initial-state
radiation in a small-$x$ event at a $p \bar{p}$ collider,  together with labels for the
kinematics.  According to the CCFM evolution equation, the emission of
partons during the initial cascade is only allowed in an
angular-ordered region of phase space. The maximum allowed angle $\Xi$
is defined by the hard scattering quark box, producing the heavy quark pair.
 In terms
of Sudakov variables the quark pair momentum is written as:
\begin{equation}
p_q + p_{\bar{q}} = \Upsilon (p^{(1)} + \Xi p^{(2)}) + Q_t
\end{equation}
where $p^{(1)}$ and $p^{(2)}$ are the four-vectors of
 incoming protons, 
respectively and $Q_t$ is the transverse momentum of the quark pair
in the laboratory frame.
Similarly, the momenta $p_i$ of the gluons emitted during the initial
state cascade are given by (here treated massless):
\begin{equation}
p_i = \upsilon_i (p^{(1)} + \xi_i p^{(2)}) + p_{ti} \;  , \;\; 
\xi_i=\frac{p_{ti}^2}{s \upsilon_i^2},
\end{equation}
with $\upsilon_i = (1 - z_i) x_{i-1}$, $x_i = z_i x_{i-1}$ and
$s=(p^{(1)}+p^{(2)})^2$ being the squared center of mass energy.
The variable $\xi_i$ is connected to the angle of the emitted gluon
with respect to the incoming proton and $x_i$ and $\upsilon_i$ are the
momentum fractions of the exchanged and emitted gluons, while $z_i$ is
the momentum fraction in the branching $(i-1) \to i$ and $p_{ti}$ is
the transverse momentum of the emitted gluon $i$.
\par
The angular-ordered region is then specified by (for the lower part of the
cascade in Fig.~\ref{CCFM_variables}, the upper part is obtained by properly
exchanging the variables):
\begin{equation}
\xi_0 < \xi_1< \cdots < \xi_n < \Xi
\end{equation}
which becomes:
\begin{equation}
z_{i-1} q_{i-1} < q_{i} 
\end{equation}
where the rescaled transverse momenta $q_{i}$ of the emitted
gluons is defined by:
\begin{equation}
 q_{i} = x_{i-1}\sqrt{s \xi_i} = \frac{p_{ti}}{1-z_i}
\end{equation}
\par
The CCFM equation for the unintegrated gluon density can be
written~\cite{CCFMd,jung_salam_2000,Salam,Martin_Sutton}
 as an integral equation:
\begin{equation}
{\cal A} (x,\kt,\Pmax ) = {\cal A}_0 (x,\kt,\Pmax ) + \int \frac{dz }{z} 
\int \frac{d^2 q}{\pi q^{2}} \Theta(\Pmax - zq) \Delta_s(\Pmax ,z q) 
\tilde{P}(z,q,\kt) {\cal A}\left(\frac{x}{z},\ktp,q\right) 
\label{CCFM_integral} 
\end{equation}  
with $\vec{k}'_{t} = | \vec{k}_{t} + (1-z) \vec{q}|$ and $\Pmax$ being
the upper scale for the last angle of the emission: $\Pmax > z_n q_n$,
$q_n > z_{n-1} q_{n-1}$, ..., $q_{1} > Q_0$. Here $q$
is used as a shorthand notation for the 2-dimensional vector of the
rescaled transverse momentum $\vec{q}\equiv\vec{q}_t=\vec{p}_t/(1-z)$.  The
splitting function $\tilde{P}(z,q,\kt)$ 
and the Sudakov form factor $\Delta_s(\Pmax ,z q)$ are given explicitly
in~\cite{jung_salam_2000}.

\subsection{The unintegrated gluon density}
In ~\cite{jung_salam_2000} 
the unintegrated gluon density $x {\cal A}(x,k_{t}^2,\Pmax)$ has been
obtained from a fit to the structure function 
$F_2(x,Q^2)$~\footnote{A Fortran program for the unintegrated gluon
density  $x {\cal A}(x,k_{t}^2,\Pmax)$ can be obtained from \cite{CASCADEMC}}. 
\par
\begin{figure}[htb] 
\vskip -0.5cm
\begin{center}
\epsfig{figure=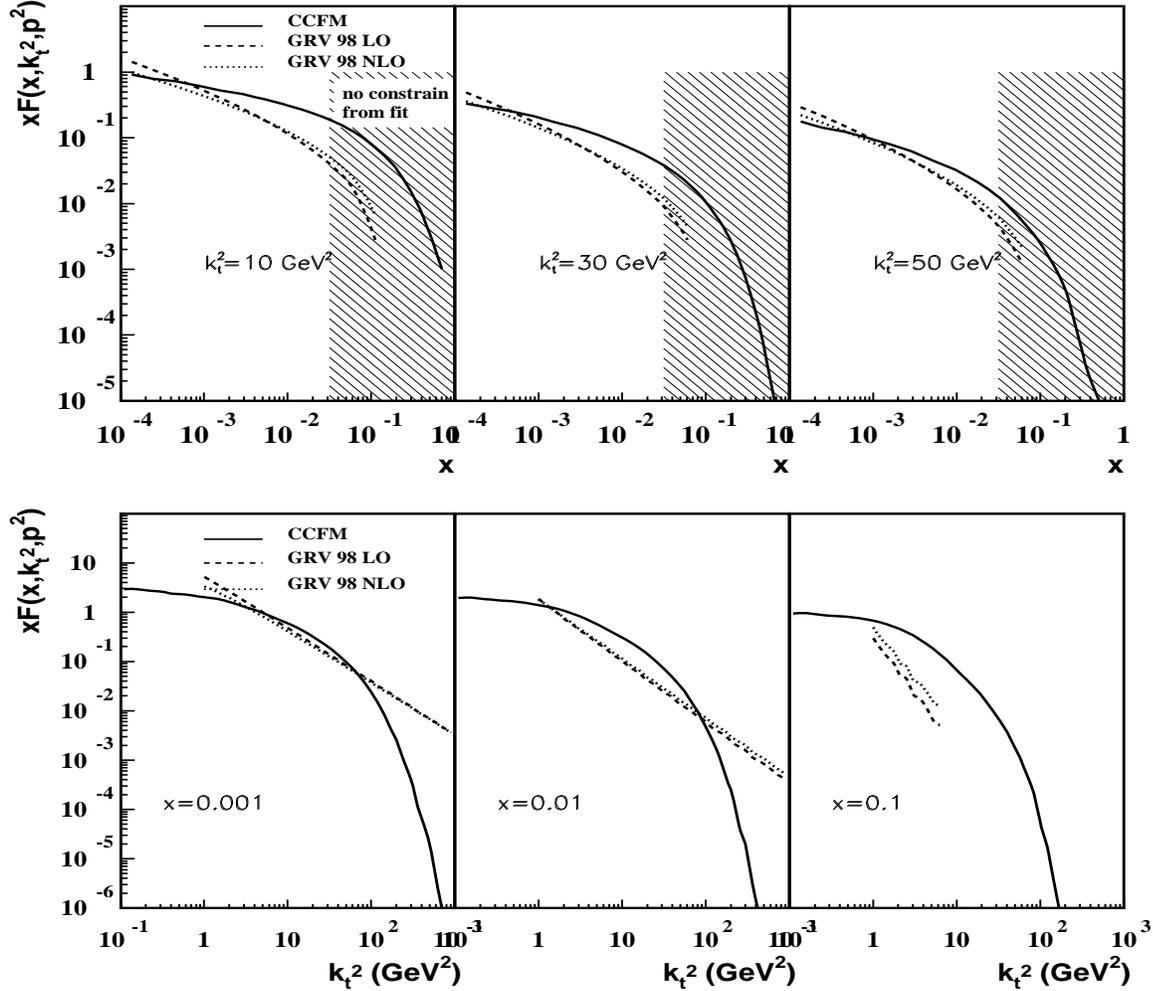,width=18cm,height=15cm} 
\caption{{\it
The  CCFM $k_t$ dependent (unintegrated) gluon density
~\protect\cite{jung_salam_2000,CASCADEMC} at $\Pmax =10$ GeV 
 as a function of
$x$ for different values of $k_t^2$ (upper)
and as a function of 
$k_t^2$ for different values of $x$ (lower) compared to
$\frac{dxG(x,\mu^2)}{d\mu^2}$  with $xG(x,\mu)$ being the
collinear gluon density of 
GRV~98~\protect\cite{GRV98} in LO and NLO. 
 }}\label{gluon} 
 \end{center}
\end{figure}

In Fig.~\ref{gluon} the CCFM unintegrated gluon density distribution 
as a function of  $x$ and  $k_t^2$ is shown and compared to
\begin{equation}
{\cal F}(x,k_t^2) \simeq \left.
\frac{d xG(x,\mu^2)}{d\mu^2}\right|_{\mu^2 = k_t^2}
\end{equation}
with $xG(x,\mu)$ being the collinear gluon density of 
GRV~98~\protect\cite{GRV98} in LO and NLO.  
\par
The unintegrated gluon density can be related to the integrated one by:
\begin{equation}
\left. xG(x,\mu) \right|_{\mu = \Pmax}
 \simeq \int_0 ^{\Pmax^2} d k_t^2 x {\cal A}(x,k_{t}^2,\Pmax)
\label{ccfm_intglu}
\end{equation}
Here the dependence on the scale of the maximum angle
$\Pmax$  is made explicit: the
evolution proceeds up to a maximum angle 
related to $\Pmax$, which plays the role of the
evolution scale in the collinear parton densities. This becomes
 obvious since 
\begin{equation}
\Pmax^2 = x_g^{(2)} \Xi s =
x_g^{(1)} x_g^{(2)} s
 = \hat{s} + Q_t^2
\end{equation}
The last expression is derived by using 
$p_Q + p_{\bar{Q}} \simeq x_g^{(2)} p_2 + x_g^{(1)} p_1 + Q_t $,
$\Xi \simeq x_g^{(1)}/x_g^{(2)}$ and 
and $\hat{s} = x_g^{(1)} x_g^{(2)} s - Q_t^2$.
This can be compared to a possible choice of 
the renormalization and factorization scale $\mu^2$  
in the collinear approach with $\mu^2 = Q_t^2 + 4 \cdot m_Q^2$ and the
similarity between $\mu$ and $\Pmax$ becomes obvious.
\par
\begin{figure}[htb] 
  \vspace*{1mm} 
  \begin{center}
\epsfig{figure=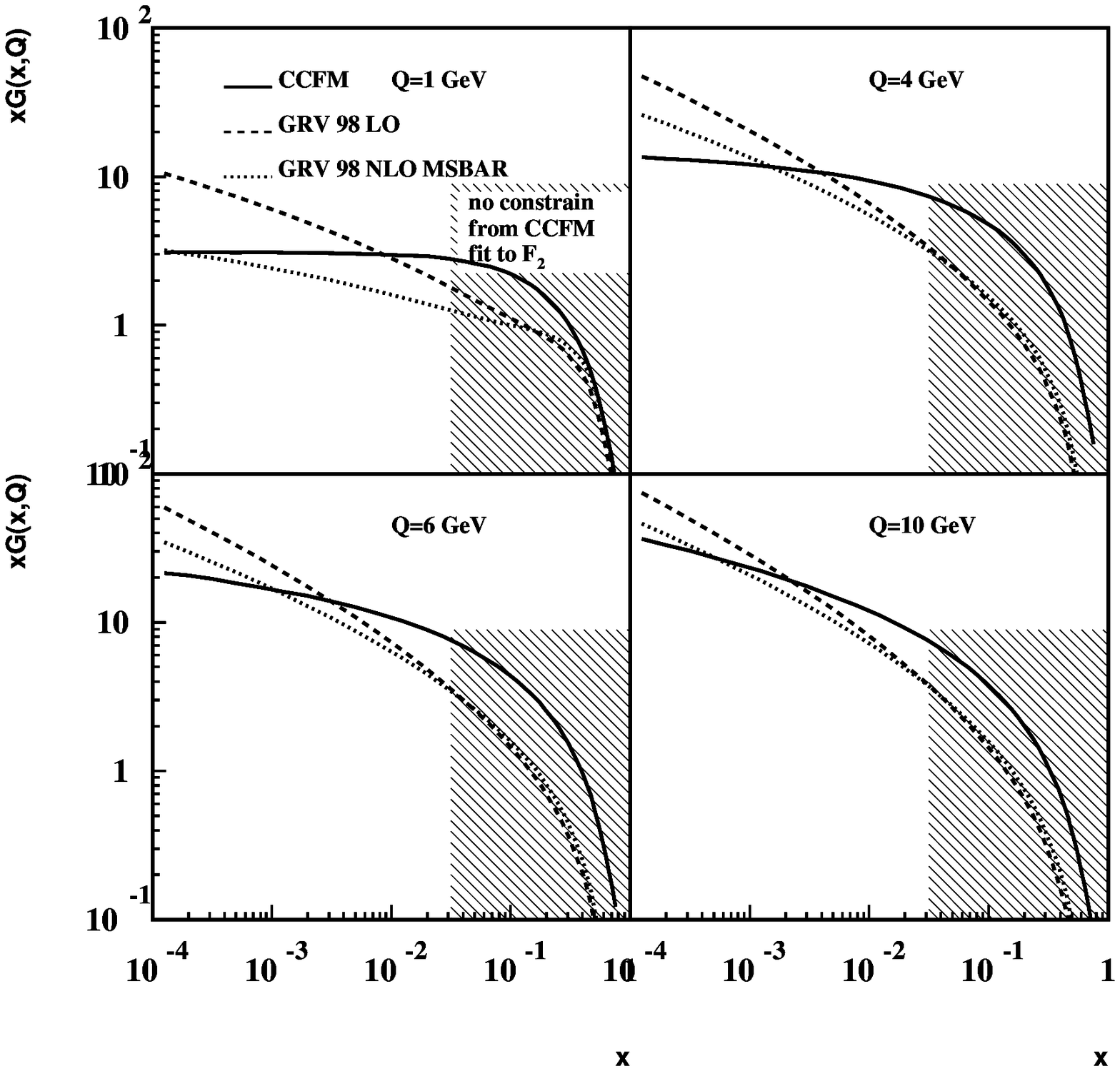,width=13cm,height=12cm} 
\caption{{\it 
The  CCFM gluon density (solid line) integrated over $k_t$
as a function of $x$ for different values of $Q$. For comparison the
GRV~98~\protect\cite{GRV98} gluon density in LO 
(dashed line) and NLO (dotted line) is also shown.
 }}\label{intglu} 
 \end{center}
\end{figure}

In Fig.~\ref{intglu} the CCFM gluon density integrated over $k_t$
according to eq.(\ref{ccfm_intglu}) is compared to
the gluon densities of GRV~98~\cite{GRV98} in LO and NLO.
It is interesting to note that the CCFM gluon density is flat for $x\to 0$ at
the input scale $Q=1$ GeV. 
Even at larger scales the collinear gluon densities rise faster with decreasing
$x$ than the CCFM gluon density. However, 
 after evolution and convolution with the off-shell matrix element 
 the scaling violations of $F_2(x,Q^2)$ and the rise of $F_2$
towards small $x$ is reproduced,
 as shown in \cite[Fig.~4 therein]{jung_salam_2000}. 
A similar trend
is observed in the collinear fixed order calculations, when going from LO to
NLO: at NLO the gluon density is less steep at small $x$, because part of the
$x$ dependence is already included in the NLO $P_{qg}$ splitting function, as
argued in \cite{catani-feb2000}.
\par
From Fig.~\ref{intglu} one can see, that the integrated gluon density from CCFM
is larger in the medium $x$ range, than the ones from the collinear approach.
Due to an additional $1/k_t^2$ suppression in the off-shell matrix elements,
the gluon density obviously has 
to be larger to still reproduce the same cross section. In 
addition  only gluon
ladders are considered in the $k_t$ - factorization approach used here,
which means that the sea quark contribution to the
structure function $F_2(x,Q^2)$ comes entirely from boson gluon fusion, without
any contribution from the intrinsic quark sea.
One also should remember, that the relation in eq.(\ref{ccfm_intglu}) is only 
approximately true, since the gluon density itself is not a 
physical observable.
\section{{\boldmath $b \bar{b}$} production at the TEVATRON}
\begin{figure}[htb]
\begin{minipage}{0.45\textwidth}
  \begin{center}
\epsfig{figure=d0bbar_add.epsi,width=8cm,height=8cm}
\caption{{\it 
Cross section for $b\bar{b}$ production with $|y^b| < 1 $ as a function of
$p_T^{min}$. Shown are the D0~\protect\cite{D0_bbar} 
data points, the fixed
order NLO prediction, and the prediction of \CASCADE.
    }}\label{d0_bbar}
    \end{center}
\end{minipage} 
\hspace*{0.5cm}
\begin{minipage}{0.45\textwidth}
\epsfig{figure=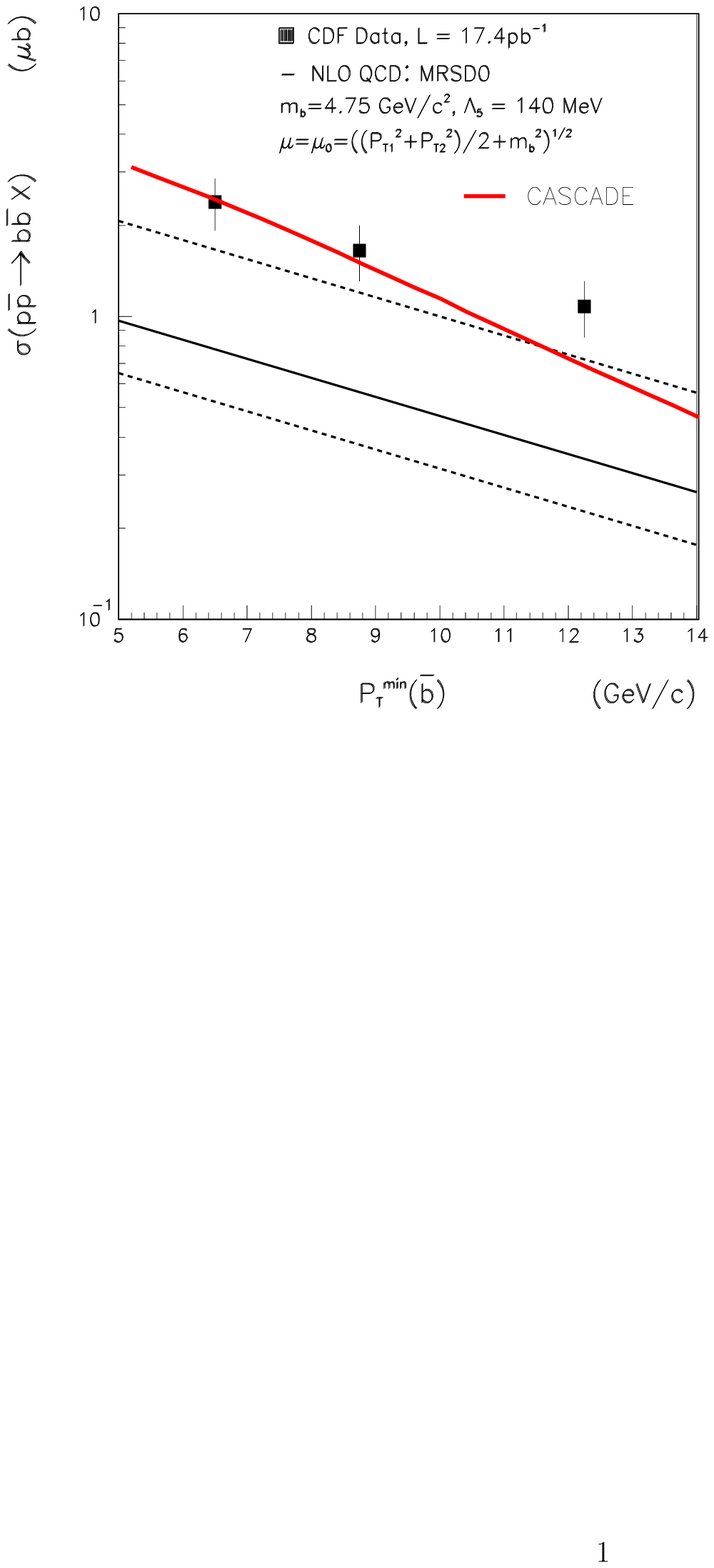,width=8cm,height=8cm}
\caption{{\it 
Cross section for $b\bar{b}$ production with $|y^b| < 1 $
and $p_T^{min}(b) > 6.5 $~GeV as a function of
$p_T^{min}(\bar{b})$. Shown are the CDF~\protect\cite{CDF_bbar} 
data points, the fixed
order NLO prediction, and the prediction of \CASCADE.
    }}\label{cdf_bbar}
\end{minipage}   
\end{figure}

The cross section for $b \bar{b}$ production in $p\bar{p}$ collision at
$\sqrt{s}=1800$~GeV is calculated with 
\CASCADE\ ~\cite{jung_salam_2000,CASCADEMC}, which is a Monte Carlo
implementation of the CCFM approach described above.
The off-shell matrix element as given
in~\cite{CCH} for heavy quarks is used with $m_b=4.75$ GeV. The scale
$\mu$ used in
$\alpha_s(\mu^2)$ is set to $\mu^2= m_T^2 = m_b^2 + p_T^2$ (as in
\cite{jung_salam_2000}), with $p_T$ being
the transverse momentum of the heavy quarks in the $p\bar{p}$ 
center-of-mass frame.
In Fig.~\ref{d0_bbar} the prediction for the cross section for $b\bar{b}$
production with pseudo-rapidity
$|y^b| < 1 $ is shown as a function of $p_T^{min}$ and
compared to the measurement of D0~\cite{D0_bbar}. Also shown is the NLO
prediction from \cite[taken from~\cite{D0_bbar}]{Frixione-Mangano}. 
In Fig.~\ref{cdf_bbar} the measured
cross section of CDF ~\cite{CDF_bbar} is shown 
for 3 values of $p_T^{min}(\bar{b}) $
with the kinematic constraint of  $|y^b|,|y^{\bar{b}}| < 1 $ and 
$p_T^{min}(b) > 6.5 $~GeV  together
with the prediction from \CASCADE\ and the NLO calculation
from~\cite[taken from~\cite{CDF_bbar}]{Mangano-Nason-Ridolfi}.
\begin{figure}[htb]
  \vspace*{2mm}
  \begin{center}
\begin{minipage}{0.45\textwidth}
  \begin{center}
\epsfig{figure=d0_ptmu.epsi,width=8cm,height=8cm}
\vskip -0.2cm
\caption{{\it 
Cross section for muons from $b$-quark decays as a function of $p_T^{\mu}$ (per
unit rapidity) as measured by  D0~\protect\cite{d0_bb_muon} compared to the
prediction of \CASCADE\ .
    }\label{d0_ptmuon}}
    \end{center}
\end{minipage} 
\hspace*{0.5cm}
\begin{minipage}{0.45\textwidth}
\epsfig{figure=d0_etamu.epsi,width=8cm,height=8cm}
\vskip -0.2cm
\caption{{\it 
Cross section for muons from $b$-quark decays as a function of $|y^{\mu}|$ 
for two different $p_T^{\mu}$ cuts 
as measured by  D0~\protect\cite{d0_bb_muon} compared to the
prediction of \CASCADE\ .
    }\label{d0_etamuon}}
\end{minipage}   
    \end{center}
\end{figure}
In all cases the NLO calculation used $m_b=4.75$ GeV and the factorization and
renormalization scales were set to $\mu^2= m_T^2 = m_b^2 + p_T^2$. Both, D0 and
CDF measurements are above the NLO predictions by a factor of $\sim 2$. The
\CASCADE\  predictions are in reasonable agreement with the measurements.
A similarly good description of the D0 and CDF data 
has been obtained in \cite{Hagler_bbar} using also $k_t$ - factorization but
supplemented with a BFKL type unintegrated gluon density.
 It is
interesting to note, that the CCFM unintegrated gluon density has been obtained
from inclusive $F_2(x,Q^2)$ at HERA. In this sense, the prediction of \CASCADE\ 
is a parameter free prediction of the $b\bar{b}$ cross section in $p\bar{p}$
collisions. This also shows for the first time 
evidence for the universality of the unintegrated CCFM gluon distribution.
\par
Since \CASCADE\ generates full hadron level events, direct comparisons with
measured cross sections 
for the production of $b$ quarks decaying semi-leptonically into muons
are possible.
The muon cross sections as a function of the transverse momentum $p_T^{\mu}$ 
and pseudo-rapidity $|y^{\mu}|$ 
as measured by D0~\cite{d0_bb_muon} are compared to the \CASCADE\ prediction in  
Fig.~\ref{d0_ptmuon} and Fig.~\ref{d0_etamuon}. Both the $p_T^{\mu}$ and 
$|y^{\mu}|$ cross sections are well described, whereas NLO
calculations underestimate the cross sections by a factor of $\sim 4$ in the
small angle range~\cite{d0_bb_muon}. 
\par
It has been argued in~\cite{sjostrand-norrbin},
that within the collinear approach supplemented with DGLAP
parton showers, heavy quark excitation in form of the processes $Qg\to Qg$ and
$Qq\to Qq$ contributes significantly to the $b\bar{b}$ cross section in the
central rapidity region.
The $b\bar{b}$ measurements at the TEVATRON 
can be reasonably well described~\cite{rfield_bbar} 
 by the price of including 
heavy quark excitation and a  
heavy quark component in the structure function of the 
proton.
However $k_t$-factorization is superior to the 
collinear approach with heavy quark excitation, 
since in the $k_t$-factorization approach 
heavy quarks are produced  perturbatively only via hard scattering matrix
elements, without further additional assumptions.
\begin{figure}[htb]
  \vspace*{2mm}
  \begin{center}
\epsfig{figure=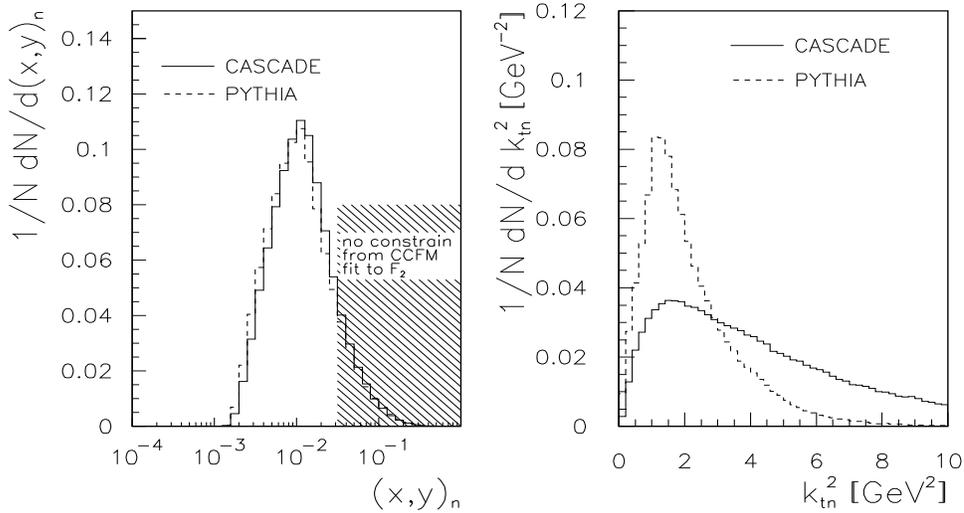,width=14cm,height=8cm}
\caption{{\it 
Comparison of $x_n$ and $k_{tn}$ distributions of the gluons entering the
process $g +g \to b + \bar{b}$. Shown are the predictions from \CASCADE\ 
representing  $k_t$-factorization with the CCFM unintegrated gluon density
and also from \PYTHIA\ representing the collinear approach supplemented with
initial and final state DGLAP parton showers to cover the phase space of $p_T$
ordered QCD cascades. 
    }}\label{kin_bbar}
    \end{center}
\end{figure}
\par
In Fig.~\ref{kin_bbar} the 
$x_n$ and $k_{tn}$ distributions  of the gluons
entering the hard scattering process 
$g +g \to b + \bar{b}$ (see Fig.~\ref{CCFM_variables}) are shown
and the predictions from the $k_t$ - factorization approach (\CASCADE\ ) 
are compared to the
standard collinear approach (here \PYTHIA\ ~\cite{\PYTHIAMC} with LO
  $gg \to  b  \bar{b}$ matrix elements 
supplemented with DGLAP parton showers to simulate higher order 
effects). Whereas the $x_n$ distributions agree reasonably well, a significant
difference is observed in the  $k_{tn}$ distribution. However this is not
surprising: 
the $k_t$ factorization approach includes
a large part of the fixed order NLO corrections (in collinear factorization).
Such corrections are $gg \to Q \bar{Q} g$, where the final state gluon 
can have any kinematically allowed transverse momentum, 
which could be regarded as a first step
toward a non-$p_T$ ordered QCD cascade.
\begin{figure}[htb]
  \vspace*{2mm}
  \begin{center}
\epsfig{figure=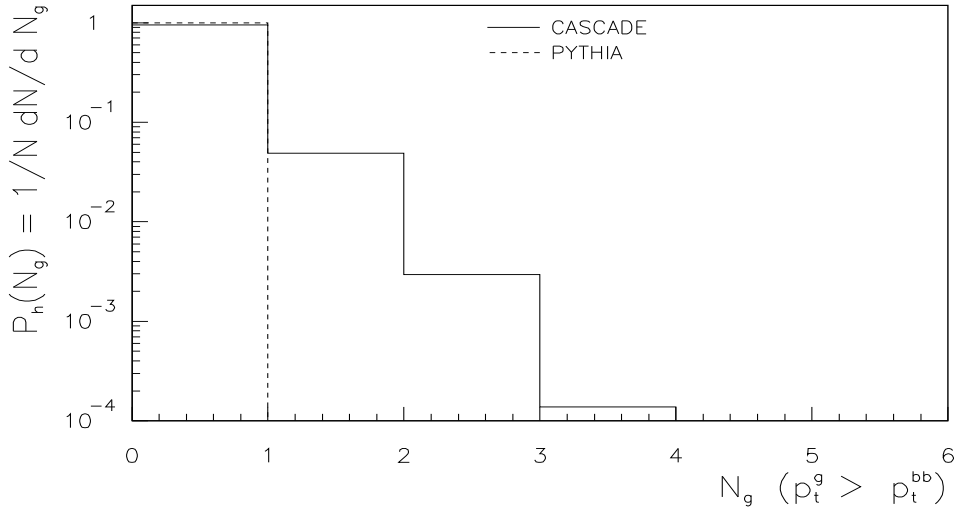,width=14cm,height=8cm}
\caption{{\it 
Probability  $P_h(N_g) = \frac{1}{N} \frac{dN}{dN_g}$ of hard gluon radiation 
with $p_{t}^{g} > m_T $ 
in $b\bar{b}$ production. Shown are calculations in the $k_t$ - factorization
approach (\CASCADE ) and from the LO DGLAP approach (\PYTHIA ).
    }}
    \label{Nglu_bbar}
    \end{center}
\end{figure}
The importance of higher order QCD radiation can be estimated from the number of
hard gluon radiation $N_g$ 
with a transverse momentum of the gluon $p_T^g > m_T$, with $m_T =
(p_T^{b}+p_T^{\bar{b}})/2 + m_b$,
by defining the probability $P_h$:
\begin{equation}
P_h(N_g) = \left. \frac{1}{N} \frac{dN}{dN_g}\right|_{p_T^g >
m_T}. 
\end{equation} 
In a leading order calculation in the collinear approach, even
supplemented with DGLAP parton showers, it is expected that
$P_h (N_g\geq 1) \to 0$, because of the approximate transverse ordering 
constraint in DGLAP. In a fixed order $\alpha_s^2$ (NLO) calculation,
$P_h (N_g= 1) \neq 0$, but $P_h (N_g \geq 2) = 0$. 
In Fig.~\ref{Nglu_bbar} the probability $P_h$ is shown as a function of $N_g$
for the production of $b\bar{b}$ events with $p_T^b > 5$ GeV, $|y^b| < 1$
obtained in the $k_t$ - factorization approach via \CASCADE. 
Also shown  for comparison is the
collinear approach in LO supplemented with DGLAP parton showers
obtained from \PYTHIA. In the $k_t$-factorization approach a significant 
contribution of hard higher order QCD radiation with 
$p_T^g >m_T$ is obtained.
\section{{\boldmath $b \bar{b}$} production at HERA}
In \cite{jung_salam_2000} the prediction of \CASCADE\
for the total $b\bar{b}$ cross section was compared to
the extrapolated measurements of the
H1~\cite{H1_bbar} and ZEUS~\cite{ZEUS_bbar} experiments at HERA. 
Since \CASCADE\ generates full hadron level events, 
a direct comparison with measurements can be done,
before extrapolating the measurement over the full phase space 
to the total $b\bar{b}$ 
cross section. ZEUS~\cite{ZEUS_bbar} has measured the dijet 
cross section which can be attributed to bottom production 
by demanding an electron 
inside one of the jets. In the kinematic range of
$Q^2 < 1$~GeV$^2$, $0.2 < y< 0.8$, at least two jets with
$E_T^{jet1(2)}>7(6)$~GeV and $|\eta^{jet}|<2.4$ and a prompt electron with
$p_T^{e^-} > 1.6$~GeV and $|\eta^{e^-}|<1.1$, 
ZEUS~\cite{ZEUS_bbar} quotes the cross section: 
\begin{equation}
\sigma^{b \to e^-}_{e^+p \to e^+ + dijet +e^- +X } =24.9 \pm 6.4^{+4.2}_{-7.3}
\mbox{ pb }\hspace{1cm}\mbox{(ZEUS~\cite{ZEUS_bbar})}
\end{equation}
with the statistical (first) and systematic (second) error given.
Within the same kinematic region and applying the same jet algorithm 
\CASCADE\  predicts:
\begin{equation}
\sigma^{b \to e^-}_{e^+p \to e^+ + dijet +e^- +X } = 20.3 ^{+1.6}_{-1.9}
\mbox{ pb,}
\end{equation}
where the error reflects the variation of $m_b=4.75\mp 0.25$~GeV. 
This value agrees with the measurement within the
statistical error.
It has been shown in~\cite{jung_ringberg2001}, that the extrapolation to the 
 $b\bar{b}$ cross section cross section 
includes large model
uncertainties and the comparison of extrapolated cross sections 
with predictions from NLO calculations are questionable. 
\begin{figure}[htb]
\vspace*{2mm}
\begin{center}
\epsfig{figure=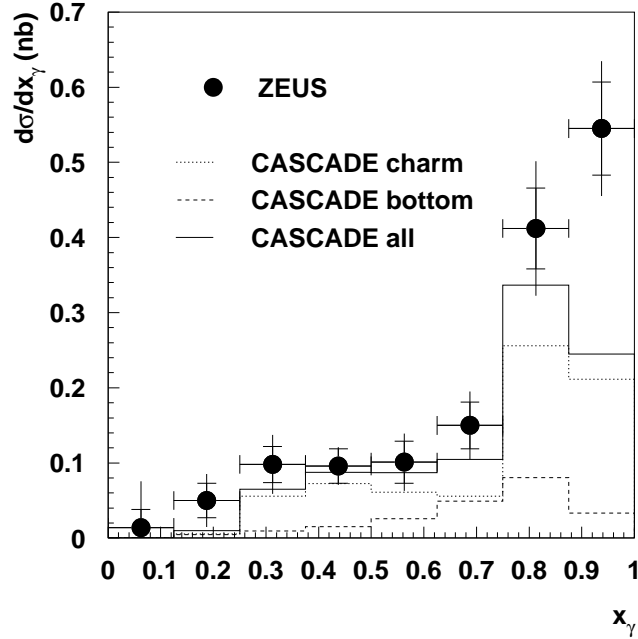,width=10cm,height=10cm}
\end{center}
\vskip -1cm
\caption{{\it The differential cross section $d\sigma/dx^{obs}_{\gamma}$
for heavy quark decays as measured by ZEUS~\protect\cite{ZEUS_bbar} and compared
to \CASCADE. Shown are the charm, bottom and the sum of both contributions to
the cross section (Note there is no additional $K$ factor applied).
}}\label{zeus_xgamma}
\end{figure}
\par
It is interesting to note, that the \CASCADE\
result agrees well with the \PYTHIA\ result
for the di-jet plus electron cross section, 
if heavy quark excitation is included in \PYTHIA .
As in the case of $b\bar{b}$ production at the TEVATRON higher order 
QCD effects are important, which are already included in the 
$k_t$ - factorization approach.
\par
In Fig.~\ref{zeus_xgamma} the differential 
cross section for heavy quark decays (charm and bottom)
as a function of $x_{\gamma}$ predicted by \CASCADE\  
is compared with the measurement of ZEUS~\cite{ZEUS_bbar}.
A significant fraction of the cross section has $x_{\gamma}<1$, 
which is 
similar to the observation in charm photo-production~\cite{jung_salam_2000}.
In LO in the collinear factorization
approach this is attributed to resolved photon processes. However, in $k_t$ -
factorization, the $x_{\gamma}$ distribution is explained 
naturally~\cite{baranov_zotov_2000}, because 
gluons in the initial state need not to be radiated 
in a $p_T$ ordered region
and therefore can give rise to a high $p_T$ jet with transverse momentum larger
than that of the heavy quarks.
\par
The prediction of \CASCADE\  has also been compared to the measurement of 
H1~\cite{H1_bbar} for electro-production cross section in 
$ Q^2 < 1$ GeV$^2$, $0.1< y < 0.8$, $p_{\perp}^{\mu} < 2$ GeV and $35^o <
\theta^{\mu} < 130^o$:
$$\sigma(ep \to e' b \bar{b}X \to \mu X')
= 0.176\pm0.016(stat.)^{+0.026}_{-0.017}(syst.)\; \mathrm{nb}$$
This cross section already includes the extrapolation from measured jets to
the muon.
In the same kinematic range \CASCADE\ predicts:
\begin{equation*}
\sigma(ep \to e' b \bar{b}X \to \mu X') = 
 0.066^{+0.009}_{-0.007} \;\mathrm{nb}.
\end{equation*} 
which is a factor $\sim 2.6 $ below the measurement. It has been shown
in~\cite{jung_ringberg2001} that the experimental results from H1 and ZEUS are
different by a factor $\sim 2$. 
It has been checked, that the difference between the experiments is not due to
different kinematics.

\section{Conclusion}
Bottom production at the TEVATRON can be reasonably well
described using the $k_t$- factorization approach with off-shell matrix elements
for the hard scattering process. 
Also the cross section for $b$ quarks decaying semi-leptonically into
muons at small angles is
reasonably well described, whereas NLO calculations fall below the data by a
factor of $\sim 4$. One essential ingredient for the satisfactory
description is the unintegrated gluon density, which was obtained by a CCFM
evolution fitted to structure function data at HERA, showing
evidence for the universality of the unintegrated gluon density.
The comparison with the data was performed with the \CASCADE\ Monte
Carlo event generator, which implements $k_t$-factorization together with the
CCFM unintegrated gluon density.
\par
Measurements of bottom production at HERA are also
compared to predictions from \CASCADE. 
The visible dijet plus electron cross section attributed to
$b$-production as measured by ZEUS could be reproduced within the statistical
error. However, the situation is different with the H1 measurement: the visible
muon cross section is already a factor of 2.6 
above the prediction from \CASCADE. 
Further measurements, also differential, are     
desirable to clarify the situation at HERA.
\par
In general the $k_t$-factorization approach has now proven to be successful 
in a wide kinematic range. It is worthwhile to note, 
that the approach presented here,
in addition to the description of $b \bar{b}$ production, is also able  
to describe charm and jet-production at HERA as well as the
inclusive structure function $F_2$. 
It is the advantage of this approach that contributions to the cross section,
which are of NLO or even NNLO nature in the collinear ansatz,
are consistently included in $k_t$-factorization
due to the off-shellness of the gluons, which enter 
into the hard scattering process.  
\section{Acknowledgments}
I have learned a lot from discussions with B.~Andersson, G.~Gustafson, 
G.~Ingelman, L.~L\"onnblad and T. Sj\"ostrand.
I am grateful to G.~Salam for all his patience and his help in all
different kinds of discussions concerning CCFM and a backward
evolution approach. Many thanks also go to 
S.~Baranov and N.~Zotov who were the first raising my interest in 
$k_t$-factorization.
I am grateful to E.~Elsen, L.~Gladilin, G.~Ingelman, L.~J\"onsson,
P.~Schleper for careful reading and comments on the manuscript. 
No words are adequate for everything I have with Antje.
I also want to thank the DESY directorate for
hospitality and support.

\end{document}